\documentclass[12pt]{article}
\usepackage{epsfig,graphicx}
\usepackage{fpcp04}

\def\be{\begin{eqnarray}}
\def\en{\end{eqnarray}}

\def\ov{\overline}

\def\CP{{\it CP}~}

\def\beq{\begin{equation}}
\def\eeq#1{\label{#1}\end{equation}}
\def\eeqn{\end{equation}}

\def\beqa{\begin{eqnarray}}
\def\eeqa#1{\label{#1}\end{eqnarray}}
\def\eeqan{\end{eqnarray}}

\let\bar=\overbar

\def\Dslash{\not{\hbox{\kern-4pt $D$}}}
\def\dslash{\not{\hbox{\kern-2pt $\del$}}}

\def\msb{{\bar{\ssstyle M \kern -1pt S}}}

\def\BB0bar{B^0 {\overline B}^0}
\def\BB0dbar{B_d^0 {\overline B}_d^0}
\def\BB0sbar{B_s^0 {\overline B}_s^0}

\RequirePackage{xspace}

\usepackage{relsize}
\def\babar{\mbox{\slshape B\kern-0.1em{\smaller A}\kern-0.1em
    B\kern-0.1em{\smaller A\kern-0.2em R}}}

\def\en         {\ensuremath{e^-}\xspace}   

\def\Kbar  {\kern 0.2em\overline{\kern -0.2em K}{}\xspace}

\def\Kz    {\ensuremath{K^0}\xspace}
\def\Kzb   {\ensuremath{\Kbar^0}\xspace}
\def\KzKzb {\ensuremath{\Kz \kern -0.16em \Kzb}\xspace}
\def\Kp    {\ensuremath{K^+}\xspace}
\def\Km    {\ensuremath{K^-}\xspace}

\def\KpKm  {\ensuremath{\Kp \kern -0.16em \Km}\xspace}

\def\Dbar    {\kern 0.2em\overline{\kern -0.2em D}{}\xspace}

\def\Dz      {\ensuremath{D^0}\xspace}
\def\Dzb     {\ensuremath{\Dbar^0}\xspace}
\def\DzDzb   {\ensuremath{\Dz {\kern -0.16em \Dzb}}\xspace}
\def\Dp      {\ensuremath{D^+}\xspace}
\def\Dm      {\ensuremath{D^-}\xspace}

\def\DpDm    {\ensuremath{\Dp {\kern -0.16em \Dm}}\xspace}

\def\Bbar    {\kern 0.18em\overline{\kern -0.18em B}{}\xspace}

\def\BB      {\ensuremath{B\Bbar}\xspace} 
\def\Bz      {\ensuremath{B^0}\xspace}
\def\Bzb     {\ensuremath{\Bbar^0}\xspace}
\def\BzBzb   {\ensuremath{\Bz {\kern -0.16em \Bzb}}\xspace}
\def\Bu      {\ensuremath{B^+}\xspace}
\def\Bub     {\ensuremath{B^-}\xspace}

\def\BpBm    {\ensuremath{\Bu {\kern -0.16em \Bub}}\xspace}

\mathchardef\Upsilon="7107
\def\Y#1S{\ensuremath{\Upsilon{(#1S)}}\xspace}

\mathchardef\Deltares="7101
\mathchardef\Xi="7104
\mathchardef\Lambda="7103
\mathchardef\Sigma="7106
\mathchardef\Omega="710A

\def\Deltabar{\kern 0.25em\overline{\kern -0.25em \Deltares}{}\xspace}
\def\Lbar{\kern 0.2em\overline{\kern -0.2em\Lambda\kern 0.05em}\kern-0.05em{}\xspace}
\def\Sigbar{\kern 0.2em\overline{\kern -0.2em \Sigma}{}\xspace}
\def\Xibar{\kern 0.2em\overline{\kern -0.2em \Xi}{}\xspace}
\def\Obar{\kern 0.2em\overline{\kern -0.2em \Omega}{}\xspace}
\def\Nbar{\kern 0.2em\overline{\kern -0.2em N}{}\xspace}
\def\Xb{\kern 0.2em\overline{\kern -0.2em X}{}\xspace}

\newcommand{\tev}{\ensuremath{\mathrm{\,Te\kern -0.1em V}}\xspace}
\newcommand{\gev}{\ensuremath{\mathrm{\,Ge\kern -0.1em V}}\xspace}
\newcommand{\mev}{\ensuremath{\mathrm{\,Me\kern -0.1em V}}\xspace}
\newcommand{\kev}{\ensuremath{\mathrm{\,ke\kern -0.1em V}}\xspace}
\newcommand{\ev}{\ensuremath{\mathrm{\,e\kern -0.1em V}}\xspace}
\newcommand{\gevc}{\ensuremath{{\mathrm{\,Ge\kern -0.1em V\!/}c}}\xspace}
\newcommand{\mevc}{\ensuremath{{\mathrm{\,Me\kern -0.1em V\!/}c}}\xspace}
\newcommand{\gevcc}{\ensuremath{{\mathrm{\,Ge\kern -0.1em V\!/}c^2}}\xspace}
\newcommand{\mevcc}{\ensuremath{{\mathrm{\,Me\kern -0.1em V\!/}c^2}}\xspace}

\def\mus  {\ensuremath{\rm \,\mus}\xspace}

\def\mus        {\ensuremath{\,\mu{\rm s}}\xspace}    

\def\to                 {\ensuremath{\rightarrow}\xspace}

\def\pep2{PEP-II}

\def\gsim{{~\raise.15em\hbox{$>$}\kern-.85em
          \lower.35em\hbox{$\sim$}~}\xspace}
\def\lsim{{~\raise.15em\hbox{$<$}\kern-.85em
          \lower.35em\hbox{$\sim$}~}\xspace}

\def\CP                {\ensuremath{C\!P}\xspace}

\xspace

\newcommand{\jprBase}        {Phys.\ Rev.\xspace}

\newcommand{\pr}        [1]  {\jprBase\ {\bf #1}}

\def\jetset74   {\mbox{\tt Jetset \hspace{-0.5em}7.\hspace{-0.2em}4}\xspace}

\begin{document}

\Title{Direct $CP$ Violation and Final State Interactions\\ in
Hadronic $B$ Decays}
\bigskip

%
\label{CSKimStart}

%
\author{ Hai-Yang Cheng\index{Cheng, H. Y.} }

%
\address{Institute of Physics, Academia Sinica\\
Taipei, Taiwan 115, ROC\\
}

\makeauthor\abstracts{ Final-state rescattering effects on the
hadronic $B$ decays and their impact on direct \CP violation are
examined. The phenomenology of the polarization anomaly in
$B\to\phi K^*$ decays is discussed. }

\section{Introduction}
Although mixing-induced $CP$ violation in $B$ decays has been
observed in the golden mode $B^0\to J/\psi K_S$ for several years,
a first confirmed observation of direct $CP$ asymmetry was not
established until recently in the charmless $B$ decays $\ov
B^0(B^0)\to K^\mp\pi^\pm$ by both BaBar \cite{BaBarKpi} and Belle
\cite{BelleKpi}. Also the combined BaBar and Belle measurements of
$\ov B^0\to\rho^\pm\pi^\mp$ imply a $3.6\sigma$ direct \CP
asymmetry in the $\rho^+\pi^-$ mode \cite{HFAG}. As for direct \CP
violation in $B^0\to\pi^+\pi^-$, a 5.2$\sigma$ effect was claimed
by Belle \cite{Bellepipi}, but it has not been confirmed by BaBar
\cite{BaBarpipi}.

Table 1 shows comparison of the model predictions of direct \CP
asymmetries with the world averages of experimental results
\cite{HFAG}. It appears that QCD factorization predictions
\cite{BBNS,BN} for direct \CP violation seem not consistent with
experiment, whereas pQCD results \cite{pQCD} are in the right
ballpark. Since the observation of direct \CP violation requires
at least two different contributing amplitudes with distinct
strong and weak phases, this means that one needs to explore the
final-state interaction (FSI) rescattering phases seriously which
are unlikely to be small possibly causing large compound
$CP$-violating partial rate asymmetries in aforementioned
charmless decay modes.

\begin{table}[h]
\caption{Comparison of pQCD and QCD factorization (QCDF)
predictions of direct \CP asymmetries (in \%) with experiment.
Also shown are the QCDF results including large weak annihilation
contributions denoted by QCDF(S4) and the FSI modifications to
QCDF predictions \cite{CCS} (see the main text for details). The
pQCD results for $\rho\pi$ modes are taken from \cite{Keum}.}
\begin{center}
 \begin{tabular}{| l r r r r r |}
\hline Modes & Expt. & pQCD & QCDF & QCDF(S4) & QCDF+FSI  \\
 \hline
 $\ov B^0\to K^-\pi^+$ & $-11\pm2$ & $-17\pm5$  &
 $4.5^{+9.1}_{-9.9}$ & $-4.1$ & $-14^{+1}_{-3}$  \\
 $\ov B^0\to\rho^+\pi^-$ & $-47^{+13}_{-14}$ & $-7.1^{+0.1}_{-0.2}$  &
 $0.6^{+11.6}_{-11.8}$ & $-12.9$ & $-43\pm11$  \\
 $\ov B^0\to\pi^+\pi^-$ & $37\pm24$ & $23\pm7$  &
 $-6.5^{+13.7}_{-13.3}$ & 10.3 & $64^{+3}_{-8}$  \\
 $\ov B^0\to\pi^0\pi^0$ & $28\pm39$ & $30\pm10$ &
 $45^{+52}_{-66}$ & $-19$ & $-30^{+1}_{-4}$ \\
 $\ov B^0\to \rho^-\pi^+$ & $-15\pm9$ & $12\pm2$ &
 $-1.5^{+8.6}_{-8.5}$ & 3.9 & $-24\pm6$ \\
 \hline
\end{tabular}
\end{center}
\end{table}

Besides the above-mentioned \CP violation, there exist several
other hints at large FSI effects in the $B$ sector \cite{CCS}. For
example, the measured branching ratio ${\cal
B}(B^0\to\pi^0\pi^0)=(1.5\pm0.3)\times 10^{-6}$ \cite{HFAG} cannot
be explained by either QCDF or pQCD and this may call for a
possible rescattering effect to induce $\pi^0\pi^0$. The QCDF
predictions for penguin-dominated modes such as $B\to
K^*\pi,~K\rho,~K\phi,~K^*\phi$ are consistently lower than the
data by a factor of 2 to 3 \cite{BN}. This large discrepancy
between theory and experiment indicates the importance of
subleading power corrections such as the annihilation topology
and/or FSI effects.

The QCD factorization approach provides a systematic study of
radiative corrections to naive factorization. While QCDF results
in the limit $m_b\to\infty$ are model independent,  power
corrections always involve endpoint divergences. For example, the
$1/m_b$ annihilation amplitude has endpoint divergences even at
twist-2 level and the hard spectator scattering diagram at twist-3
order is power suppressed and posses soft and collinear
divergences arising from the soft spectator quark. Since the
treatment of endpoint divergences is model dependent, subleading
power corrections generally can be studied only in a
phenomenological way. While the endpoint divergence is regulated
in the pQCD approach by introducing the parton's transverse
momentum \cite{pQCD}, it is parameterized in QCD factorization as
 \be
 X_A\equiv \int^1_0{dy\over y}={\rm ln}{m_B\over
 \Lambda_h}(1+\rho_A e^{i\phi_A}),
 \end{eqnarray}
with $\Lambda_h$ being a typical scale of order 500 MeV.

Just like the pQCD approach where the annihilation topology plays
an essential role for producing sizable strong phases and for
explaining the penguin-dominated $VP$ modes, it has been suggested
in \cite{BN} that a favorable scenario (denoted as S4) for
accommodating the observed penguin-dominated $B\to PV$ decays and
the measured sign of direct \CP asymmetry in $\ov B^0\to K^-\pi^+$
is to have a large annihilation contribution by choosing
$\rho_A=1$, $\phi_A=-55^\circ$ for $PP$, $\phi_A=-20^\circ$ for
$PV$ and $\phi_A=-70^\circ$ for $VP$ modes. The resultant direct
\CP asymmetries are shown in Table I. The sign of $\phi_A$ is
chosen so that the direct \CP violation $A_{K^-\pi^+}$ agrees with
the data. However, the origin of these phases is unknown and their
signs are not predicted. Moreover, the annihilation topologies do
not help enhance the $\pi^0\pi^0$ and $\rho^0\pi^0$ modes. As
stressed in \cite{BN}, one would wish to have an explanation of
the data without invoking weak annihilation. Therefore, it is of
great importance to study final-state rescattering effects on
decay rates and \CP violation.

\section{Final State Interactions in Hadronic $B$ decays}
In QCDF there are two hard strong phases: one from the absorptive
part of the penguin graph in $b\to s(d)$ transitions \cite{BSS}
and the other from the vertex corrections. However, these
perturbative strong phases do not lead to the correct sign of
direct \CP asymmetries observed in $K^-\pi^+$, $\rho^+\pi^-$ and
$\pi^+\pi^-$ modes. Therefore, one has to consider the
nonpertrubative strong phases induced from power suppressed
contributions such as FSIs. Based on the Regge approach, Donoghue
{\it et al.} \cite{Donoghue} have reached the interesting
conclusion that FSIs do not disappear even in the heavy quark
limit and soft FSI phases are dominated by inelastic scattering,
contrary to the common wisdom. However, it was later pointed out
by Beneke {\it et al.} \cite{BBNS} within the framework of QCD
factorization that the above conclusion holds only for individual
rescattering amplitudes. When summing over all possible
intermediate states, there exist systematic cancellations in the
heavy quark limit so that the strong phases must vanish in the
limit of $m_b\to\infty$. Consequently, the FSI phase is generally
of order ${\cal O}(\alpha_s,\Lambda_{\rm QCD}/m_b)$. In reality,
because the $b$ quark mass is not very large and far from being
infinity, the aforementioned cancellation may not occur or may not
be very effective for the finite $B$ mass. Hence, the strong phase
arising from power corrections can be in principle very sizable.

At the quark level, final-state rescattering can occur through
quark exchange and quark annihilation. In practice, it is
extremely difficult to calculate the FSI effects, but it may
become amenable at the hadron level where FSIs manifest as the
rescattering processes with $s$-channel resonances and one
particle exchange in the $t$-channel. In contrast to $D$ decays,
the $s$-channel resonant FSIs in $B$ decays is expected to be
suppressed relative to the rescattering effect arising from quark
exchange owing to the lack of the existence of resonances at
energies close to the $B$ meson mass. Therefore, we will model
FSIs as rescattering processes of some intermediate two-body
states with one particle exchange in the $t$-channel and compute
the absorptive part via the optical theorem \cite{CCS}.

The approach of modelling FSIs as soft rescattering processes of
intermediate two-body states has been criticized on several
grounds \cite{BBNS}. For example, there are many more intermediate
multi-body channels in $B$ decays and systematic cancellations
among them are predicted to occur in the heavy quark limit. This
effect of cancellation will be missed if only a few intermediate
states are taken into account. As mentioned before, the
cancellation may not occur or may not be very effective as the $B$
meson is not infinitely heavy. Hence, we may assume that two-body
$\rightleftharpoons$ $n$-body rescatterings are negligible either
justified from the $1/N_c$ argument or suppressed by large
cancellations. Indeed, it has been even conjectured that the
absorptive part of long-distance rescattering is dominated by
two-body intermediate states, while the dispersive part is
governed by multi-body states \cite{Suzuki}. At any rate, we view
our treatment of the two-body hadronic model for FSIs as a working
tool. We work out the consequences of this tool to see if it is
empirically working. If it turns out to be successful, then it
will imply the possible dominance of intermediate two-body
contributions.

The calculations of hadronic diagrams for FSIs involve many
theoretical uncertainties. Since the particle exchanged in the $t$
channel is off shell and since final state particles are hard,
form factors or cutoffs must be introduced to the strong vertices
to render the calculation meaningful in perturbation theory. We
shall parametrize the off-shell effect of the exchanged particle
as
 \be
 F(t,m)=\,\left({\Lambda^2-m^2\over \Lambda^2-t}\right)^n,
 \end{eqnarray}
normalized to unity at $t=m^2$ with $m$ being the mass of the
exchanged particle. The monopole behavior of the form factor (i.e.
$n=1$) is preferred as it is consistent with the QCD sum rule
expectation \cite{Gortchakov}. For the cutoff $\Lambda$, it should
be not far from the physical mass of the exchanged particle. To be
specific, we write $\Lambda=m_{\rm exc}+r\Lambda_{\rm QCD}$ where
the parameter $r$ is expected to be of order unity and it depends
not only on the exchanged particle but also on the external
particles involved in the strong-interaction vertex. As we do not
have first-principles calculations for form factors, we shall use
the measured decay rates to fix the unknown cutoff parameters and
then use them to predict direct \CP violation. We discuss some
applications below.

\vskip 0.2cm \noindent {\bf 2.1~~ Penguin dominated modes} \vskip
0.2cm
 Penguin dominated modes such as $B\to K\pi,~K^*\pi,~K\rho,~\phi
K^{(*)}$ receive sizable contributions from rescattering of charm
intermediate states (i.e. the so-called long-distance charming
penguins). For example, the branching ratios of $B\to \phi K$ and
$\phi K^*$ can be enhanced from $\sim 5\times 10^{-6}$  predicted
by QCDF to the level of $1\times 10^{-5}$ by FSIs via rescattering
of charm intermediate states \cite{CCS}.

\vskip 0.2cm \noindent {\bf 2.2~~ Tree dominated modes} \vskip
0.2cm \underline{$B\to D\pi$ decays}~~~The color-suppressed modes
$\ov B^0\to D^{(*)0}\pi^0$ have been measured by  Belle, CLEO and
BaBar. Their branching ratios are all significantly larger than
theoretical expectations based on naive factorization. When
combined with the color-allowed $\ov B\to D^{(*)}\pi$ decays, it
indicates non-vanishing relative strong phases among various $\ov
B\to D^{(*)}\pi$ decay amplitudes. Neglecting the $W$-exchange
contribution, a direct fit to the $D\pi$ data requires that
$a_2/a_1\approx (0.45-0.65)e^{\pm i60^\circ}$ \cite{ChengBDpi}.
The question is then why the magnitude and phase of $a_2/a_1$ are
so different from the model expectation. We found that the
rescattering from $B\to\{ D\pi,D^*\rho\}\to D\pi$ contributes to
the color-suppressed $W$-emission and $W$-exchange topologies and
accounts for the observed enhancement of the $D^0\pi^0$ mode
without arbitrarily assigning the ratio of $a_2/a_1$ a large
magnitude and strong phase as done in many previous works. Note
that the color-allowed $B\to D\pi$ decays are almost not affected
by final-state rescattering.\\

\underline{$B\to\rho\pi$ decays}~~~The color-suppressed $\rho^0
\pi^0$ mode is slightly enhanced by rescattering effects to the
order of $1.3\times 10^{-6}$, which is consistent with the
weighted average $(1.9\pm1.2)\times 10^{-6}$ of the experimental
values. However, it is important to clarify the discrepancy
between BaBar and Belle measurements for this mode, namely,
$(1.4\pm0.7)\times 10^{-6}$ \cite{BaBarrho0pi0} vs
$(5.1\pm1.8)\times 10^{-6}$ \cite{Bellerho0pi0}. Note that the
branching ratio of $\rho^0\pi^0$ is predicted to be of order
$0.2\times 10^{-6}$ in the pQCD approach \cite{Lu}, which is too
small compared to experiment as the annihilation contribution does
not help enhance its rate.\\

\underline{$B\to\pi\pi$ decays}~~~There are some subtleties in
describing $B^0\to \pi\pi$ decays. The rescattering charming
penguins in $\pi\pi$ are suppressed relative to that in $K\pi$
modes as the former are Cabibbo suppressed. Consequently, charming
penguins are not adequate to explain the $\pi\pi$ data: the
predicted $\pi^+\pi^-$ ($\sim 9\times 10^{-6}$) is too large
whereas $\pi^0\pi^0$ ($\sim 0.4\times 10^{-6}$) is too small. This
means that a dispersive long-distance contribution is needed to
interfere destructively with $\pi^+\pi^-$ so that $\pi^+\pi^-$
will be suppressed while $\pi^0\pi^0$ will get enhanced. This
contribution cannot arise from the charming penguins or otherwise
it will also contribute to $K\pi$ significantly and destroy all
the nice predictions for $K\pi$. In the topological diagrammatic
approach \cite{Chau}, this dispersive term comes from the
so-called vertical $W$-loop diagram $V$ in which the meson
annihilation topology such as $D^+ D^-\to\pi\pi$ occurs.

It is often claimed in the literature that one needs large $C/T$
to accommodate the $\pi\pi$ data, where $C$ and $T$ are
color-allowed and color-suppressed topologies, respectively. In
doing the fit to the $\pi\pi$ data, one actually redefines the
quantities $T,C,P$ to absorb other topological amplitudes, e.g.
$W$-exchange $E$, electroweak penguin $P_{\rm EW}$. Hence, a large
$C_{\rm eff}/T_{\rm eff}$ does not necessarily imply large $C/T$.
In our calculation we found $|C_{\rm eff}/T_{\rm eff}|=0.71$ for
$C_{\rm eff}=C-E-V$ and $T=T+E+V$, while $|C/T|_{\rm SD}=0.23$ and
$|C/T|_{\rm SD+LD}=0.33$ \cite{CCS}. The point is that one needs
large $V/T=0.56{\rm exp}[i72^\circ]$ to understand the $\pi\pi$
data.

\vskip 0.2cm \noindent {\bf 2.3~~ Direct \CP asymmetries} \vskip
0.2cm The strong phases in charmless $B$ decays are governed by
final-state rescattering. We see from the last column of Table 1
that direct $CP$-violating partial rate asymmetries in $K^-\pi^+$,
$\rho^+\pi^-$ and $\pi^+\pi^-$ modes are significantly affected by
final-state rescattering and their signs are different from that
predicted by the short-distance QCDF approach. The direct \CP
asymmetries in $\ov B^0\to\pi^0\pi^0,\rho^-\pi^+$ decays are also
shown in Table 1 where we see that the predictions of pQCD and
QCDF supplemented with FSIs are opposite in sign. It will be
interesting to measure direct \CP violation in these two decays to
test different models.

\section{Polarization Anomaly in $B\to\phi K^*$}
For $B\to V_1V_2$ decays with $V$ being a light vector meson, it
is expected that they are dominated by longitudinal polarization
states and respect the scaling law: $1-f_L={\cal O}(m_V^2/m_B^2)$.
However, a low value of the longitudinal fraction $f_L\approx
50\%$ in $\phi K^*$ decays was observed by both BaBar
\cite{BaBarVV} and Belle \cite{BelleVV} (see Table 2). This
polarization anomaly poses an interesting challenge for any
theoretical interpretation.

\begin{table}[t]    \label{tab:BphiKV}
 \caption{ Experimental data for
\CP averaged branching ratios (in units of $10^{-6}$) and
polarization fractions for $B\to\phi K^*$ and $\rho K^*$
\cite{BaBarVV,BelleVV}. }
\begin{center}
\begin{tabular}{c c c  c}
Mode & BaBar & Belle & Average  \\
\hline
 $f_L(\phi K^{*0})$
              & $0.52\pm0.05\pm0.02$
              & $0.52\pm0.07\pm0.05$
              & $0.52\pm0.05$
              \\
 $f_\bot(\phi K^{*0})$
              & $0.22\pm0.05\pm0.02$
              & $0.30\pm0.07\pm0.03$
              & $0.25\pm0.04$
              \\
 $f_L(\phi K^{*+})$
              & $0.46\pm0.12\pm0.03$
              & $0.49\pm0.13\pm0.05$
              & $0.47\pm0.09$
              \\
 $f_\bot(\phi K^{*+})$
              &
              & $0.12^{+0.11}_{-0.08}\pm0.03$
              & $0.12^{+0.11}_{-0.09}$
              \\

 \hline
 $f_L(\rho^0 K^{*+})$
              & $0.96^{+0.04}_{-0.15}\pm0.04$
              &
              & $0.96^{+0.06}_{-0.16}$
              \\
 $f_L(\rho^+ K^{*0})$
              & $0.79\pm0.08\pm0.04\pm0.02$
              & $0.50\pm0.19^{+0.05}_{-0.07}$
              & $0.74\pm0.08$
              \\
 \hline
\end{tabular}
\end{center}
\end{table}

Working in the context of QCD factorization, Kagan \cite{Kagan}
has argued that the lower value of the longitudinal polarization
fraction and the large transverse rate can be accommodated by the
$(S-P)(S+P)$ penguin-induced annihilation contributions. This is
so because the transverse polarization amplitude induced from the
above annihilation topologies is of the same $1/m_b$ order as the
longitudinal one. Moreover, although the penguin-induced
annihilation contribution is formally $1/m_b^2$ suppressed, it can
be ${\cal O}(1)$ numerically. Kagan showed that a fit to the data
of $f_L(\phi K^{*0})$ and $f_L(\rho^+K^{*0})$ favors $\rho_A\sim
0.5$. An alternative suggestion for the solution of the $\phi K^*$
anomaly was advocated in \cite{Hou04} that a energetic transverse
gluon from the $b\to s g$ chromodipole operator keeps most of its
quantum numbers except color when it somehow penetrates through
the $B$ meson surface and descends to a transversely polarized
$\phi$ meson. Sizable transverse components of the $B\to \phi K^*$
decay can be accommodated by having $f_\parallel>f_\perp$. Since
the gluon is a flavor singlet, this mechanism can distinguish
$\phi$ from $\rho$, hence it affects $B\to\phi K^{*},\omega K^{*}$
but not $B\to K^{*}\rho$.

Since the scaling law is valid only at short distances, one can
also try to circumvent it by considering the long-distance
rescattering contributions from intermediate states
$D^{(*)}D_s^{(*)}$ \cite{CCS,Colangelo,Ladisa}. The large
transverse polarization induced from $B\to D^*D_s^*$ will be
propagated to $\phi K^*$ via FSI rescattering. Furthermore,
rescattering from $B\to D^*D_s$ or $B\to DD^*_s$ will contribute
only to the $A_\bot$ amplitude. Recently, we have studied FSI
effects on $B\to VV$. While the longitudinal polarization fraction
can be reduced significantly from short-distance predictions due
to such FSI effects, no sizable perpendicular polarization is
found owing mainly to the large cancellations occurring in the
processes $B\to D_s^* D\to\phi K^*$ and $B\to D_s D^*\to\phi K^*$
and this can be understood as a consequence of \CP and SU(3)
symmetry. Our result is different from a recent similar study in
\cite{Colangelo}. To fully account for the polarization anomaly
(especially the perpendicular polarization) observed in $B\to \phi
K^*$, FSI from other states or other mechanism, e.g. the
aforementioned penguin-induced annihilation, may have to be
invoked.

Li \cite{LiphiKV} pointed out an interesting observation that the
polarization anomaly may be resolved in the pQCD approach provided
that the form factor $A_0^{BK^*}(0)$ of order 0.30 is employed.
This form factor is indeed very close to the result of 0.31
obtained in the covariant light-front quark model \cite{CCH}.
Since the pQCD approach tends to predict a large value of
$A_0^{BK^*}(0)$ (0.46 and 0.41 in the previous pQCD calculations
\cite{Lu,Chen}), the task is to see if a small value of $A_0$ can
be naturally produced in such an approach rather than put by hand
artificially.

As for the transverse polarization in $B\to\rho K^*$ decays, both
final-state rescattering and large annihilation scenarios lead to
$f_L(\rho K^*)\sim 60\%$, whereas $f_L(\rho K^*)\sim 1$ in the
model of \cite{Hou04}. However, none of the aforementioned models
can explain the observed disparity between $f_L(\rho^+ K^{*0})$
and $f_L(\rho^0 K^{*+})$ (see, however, \cite{LiVV} for a possible
solution). This should be clarified both experimentally and
theoretically.

\section{Acknowledgments}

I am grateful to Chun-Khiang Chua and Amarjit Soni for very
fruitful collaboration and to C.H. Chen, H.n. Li and Y.Y. Keum for
helpful discussions.

\end{document}